\documentclass[12 pt]{article}
\usepackage[utf8]{inputenc}
\usepackage{amsmath}
\usepackage{amsfonts}
\usepackage{amssymb,amsthm,bm,multirow}
\usepackage{graphicx}
\usepackage[margin=1.5in]{geometry}
\usepackage{setspace}
\usepackage{color}
\usepackage[colorlinks,bookmarksopen,bookmarksnumbered,citecolor=red,urlcolor=red]{hyperref}
\usepackage[labelformat=simple]{subcaption}

\usepackage[norelsize, linesnumbered, ruled, lined, boxed, commentsnumbered]{algorithm2e}

\SetKwBlock{Repeat}{repeat}{end}

\DeclareMathOperator*{\argmin}{arg\,min}

\title{
	\textbf{Machine Learning Based Analysis and Quantification of Potential Power Gain from Passive Device Installation}\\[.5em]
	{\large Methodology companion to \texttt{R} package \texttt{gainML}}\\
}
\author{Hoon Hwangbo$^*$, Yu Ding$^\dag$, Daniel Cabezon$^\#$\vspace{12 pt}\\ 
\footnotesize $^*$University of Tennessee, Knoxville, Tennessee, USA\\
\footnotesize $^\dag$Texas A\&M University, College Station, Texas, USA\\
\footnotesize $^\#$EDP Renov\'{a}veis, Madrid, Spain}

\date{\today}

\begin{document}
\maketitle

\section{Introduction}
This documents describes the notations, procedures and meaning of outcomes associated with the \texttt{R} package \texttt{gainML}, which is to quantify potential power gain resulting from installing passive devices on wind turbine generators (WTG). The method described here has been applied to several cases of vortex generator (VG) installation but is generally applicable to the cases of other passive device installation.

A na\"{\i}ve approach for the gain quantification is to compare the power production of a WTG before and after the passive device installation, referred to as Period~1 and Period~2, respectively. Recall that wind power production depends on various environmental factors, such as wind speed. The temporal variation of these factors contributes to power production difference between Period~1 and~2. As such, the difference in power production associated with the two periods does not simply confirm a potential power gain resulting from the passive device installation.  For an accurate quantification, the effect of passive device installation needs to be separated from the effect of environmental factors. Therefore, we build a machine learning model to control for the environmental effect and then quantify a potential gain in power production once the conditions between the two periods have been made comparable.

This machine learning-based model, when applied to the reference turbine alone, may work well if one can perfectly control for the environmental conditions. However, a perfect control for the environmental condition is extremely difficult, if at all possible, because one may not be able to identify or measure every single environmental factor that potentially affects the power production of a WTG.  In our investigation, we use a three-turbine group to add additional layers of calibration to correct for the effect of unknown or unobservable factors that may have drifted over time. The three turbines in a group are known as: (a) the reference or test turbine, the one that underwent an upgrade, (b) the baseline control turbine, a turbine nearby the test turbine that was not upgraded but provides a datum for calibrating the gain, and (c) the neutral turbine, a control turbine nearby the test turbine that was not upgraded but supplies the measurements of covariates (inputs used in the machine learning model) unaffected by the installation of the passive device.

Under this setup, we build two machine learning models, one for the test turbine and another for the control turbine, both use the covariates supplied by the neutral turbine. An underlying assumption for this calibration is that unknown or unobservable factors will have the same effect on the power production of a test turbine and a control turbine at a given time, which sounds reasonable.

While considering that, we conduct Period 1 analysis and Period 2 analysis using the machine learning models. The purpose of the Period~1 analysis is to reduce the error of a machine learning model for a given test turbine by determining a best set of environmental variables and to choose a control turbine that satisfies the assumption that the unknown or unobservable effect for the control turbine is comparable to that for the test turbine. The purpose of the Period 2 analysis is to take all decisions made in Period 1 analysis and quantifies potential power gain while applying a second calibration procedure.

In the remainder of this documentation, we describe a machine learning model that captures the effect of environmental factors in Section~\ref{sc:adpKer}, explain the details of Period 1 analysis in Section~\ref{sc:per1}, and state the step-by-step procedures of Period 2 analysis in Section~\ref{sc:per2}.

\section{Core machine learning model: an adaptive kernel regression}\label{sc:adpKer}
Suppose that $y$ refers to the power output of a WTG and $\bold{x} = (x_1, x_2, \ldots, x_p)$ denotes a set of $p$ environmental variables. Then, we would like to determine a function $f$ that defines the relationship between $y$ and $\bold{x}$ (can be considered as a high-dimensional extension of power curve) in the following model.
\begin{equation}
y_i = f(\bold{x}_i) + \epsilon_i = f(x_{1i},x_{2i}, \ldots, x_{pi}) + \epsilon_i, \quad i = 1, \ldots, n,
\end{equation}
where $i$ is the index of observations and $n$ is the total number of observations. $\epsilon_i$ models random noise, the portion of $y$ that is not explained by given $\bold{x}$; assumed to be independent and identically distributed.

We estimate $f$ by using a kernel regression estimator, known as the Nadaraya-Watson estimator, defined by
\begin{equation}
\hat{f}(\bold{x}) = \sum_{i=1}^n{\omega_i(\bold{x}) \cdot y_i},
\label{eq:est}
\end{equation}
where $\hat{f}$ is the estimator of $f$ and $\omega_i$ is a weight assigned to the $i$th observation given $\bold{x}$. Note that $\bold{x}$ does not have any subscript, meaning that $\bold{x}$ can be any value, e.g., a future realization, although it can also be an observed $\bold{x}_i$. The weight $\omega_i(\bold{x})$ is determined by a Gaussian kernel and $k$-nearest neighbor distance as
\begin{equation}
\omega_i(\bold{x}) = \frac{\mathcal{K}(\lVert \bold{x} - \bold{x}_i \rVert / R_\bold{x})}{\sum_{i=1}^n{\mathcal{K}(\lVert \bold{x} - \bold{x}_i \rVert / R_\bold{x})}}.
\end{equation}
$\lVert \cdot \rVert$ calculates the Euclidean norm, and $\mathcal{K}(\cdot)$ is a standard normal density, i.e., $\mathcal{K}(a) = \exp(-a^2/2)/\sqrt{2 \pi}$. The bandwidth parameter $R_\bold{x}$ is not a constant, but it varies with $\bold{x}$ modeling an adaptive bandwidth. We define
\begin{equation}
R_\bold{x} = d_\bold{x}(k) / 3
\label{eq:bw}
\end{equation}
where $d_\bold{x}(k)$ is the $k$th nearest distance among all the distances from $\bold{x}$ to $\bold{x}_i$ for $\forall i = 1, \ldots, n$. This adaptive bandwidth allows a smaller bandwidth where there are more data points and a larger bandwidth where there are less data points. In other words, the functional fit $\hat{f}$ would follow a local trend more if the region around an evaluation point, $\bold{x}$, is dense and would follow a global trend more if the region is sparse; this improves prediction accuracy compared to the case of using a constant bandwidth.

From the definition of the estimator in Eq.~\eqref{eq:est}--\eqref{eq:bw}, all values are given or can be calculated by using given values except the value of $k$. $k$ is the only parameter that needs to be set, and the value of $k$ will dictate the goodness of the estimator. We select $k$ that minimizes the generalized cross-validation (GCV) criterion while using only training data, $(\bold{x}_i, y_i)$ for $i=1, \ldots, n$. By fitting the model in Eq.~\eqref{eq:est} to the training data, we have
\begin{equation}
\hat{\bold{f}} = \bold{M}_n(k) \cdot \bold{y},
\end{equation}
where $\hat{\bold{f}} = \left( \hat{f}(\bold{x}_1), \hat{f}(\bold{x}_2), \ldots, \hat{f}(\bold{x}_n) \right)^\prime$ is an $n$-dimensional vector, $\bold{M}_n$ is an $n\times n$ matrix with its $(a, b)$th element given by $\omega_b(\bold{x}_a)$, and $\bold{y} = \left( y_1, y_2, \ldots, y_n \right)^\prime$ is also an $n$-dimensional vector. The parentheses following $\bold{M}_n$ signify $\bold{M}_n$'s dependence on $k$. Now, GCV criterion is defined by
\begin{equation}
\text{GCV}(k) = \frac{n^{-1} {\lVert (I - \bold{M}_n(k) ) \bold{y} \rVert}^2}{\left( n^{-1} \, \text{tr} (I - \bold{M}_n(k)) \right)^2},
\end{equation}
where $I$ is an $n\times n$ identity matrix. Then, we choose $k$ that minimizes $\text{GCV}(k)$, i.e., $k^* = \argmin_k{\text{GCV}(k)}$.

\section{Period 1 analysis}\label{sc:per1}
In Period~1 analysis, while using the machine learning model described in Section~\ref{sc:adpKer}, we determine the best set of environmental variables modeling the power production of a test turbine and choose a control turbine for which the prediction error exhibits a similar trend with that for the test turbine. In Period~1 analysis, we are interested in establishing a machine learning model that represents the power production of a WTG \emph{without any device installation} as accurately as possible; for this reason, this analysis is based on Period~1 data only.

\begin{figure}[b!]
	\centering
	\includegraphics[width=0.85\textwidth]{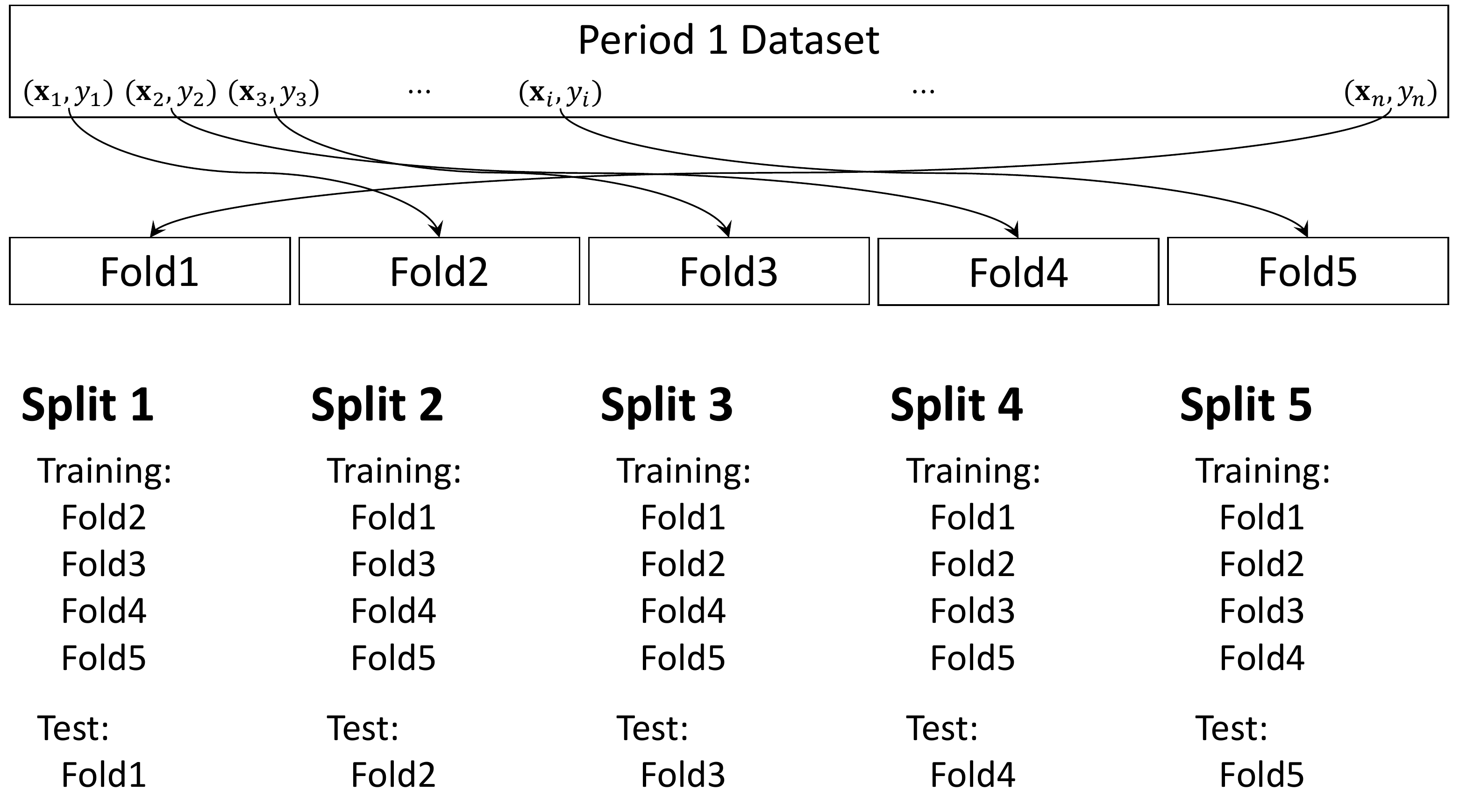}
	\caption{Data usage of Period 1 data for 5-fold CV}
	\label{fg:5fCV}
\end{figure}

For an effective usage of data, we apply 5-fold cross validation (CV) for evaluating out-of-sample prediction. 5-fold CV partitions an entire dataset (Period~1 data) into five mutually exclusive subsets of data, each of which takes 20\% of the entire dataset; the assignment of a data point to a specific subset is performed randomly. The idea is to take four folds (subsets) of data for training and the remaining single fold of data for testing, yielding 80:20 training/test data splitting. Each of the five folds can be used for testing one at a time, and this allows five times testing (and training) of a model. The multiple testing improves the estimation of out-of-sample prediction error relative to the case of a single training and testing. Figure~\ref{fg:5fCV} illustrates a data usage for the 5-fold CV.

\begin{figure}[t]
	\centering
	\includegraphics[width=0.6\textwidth]{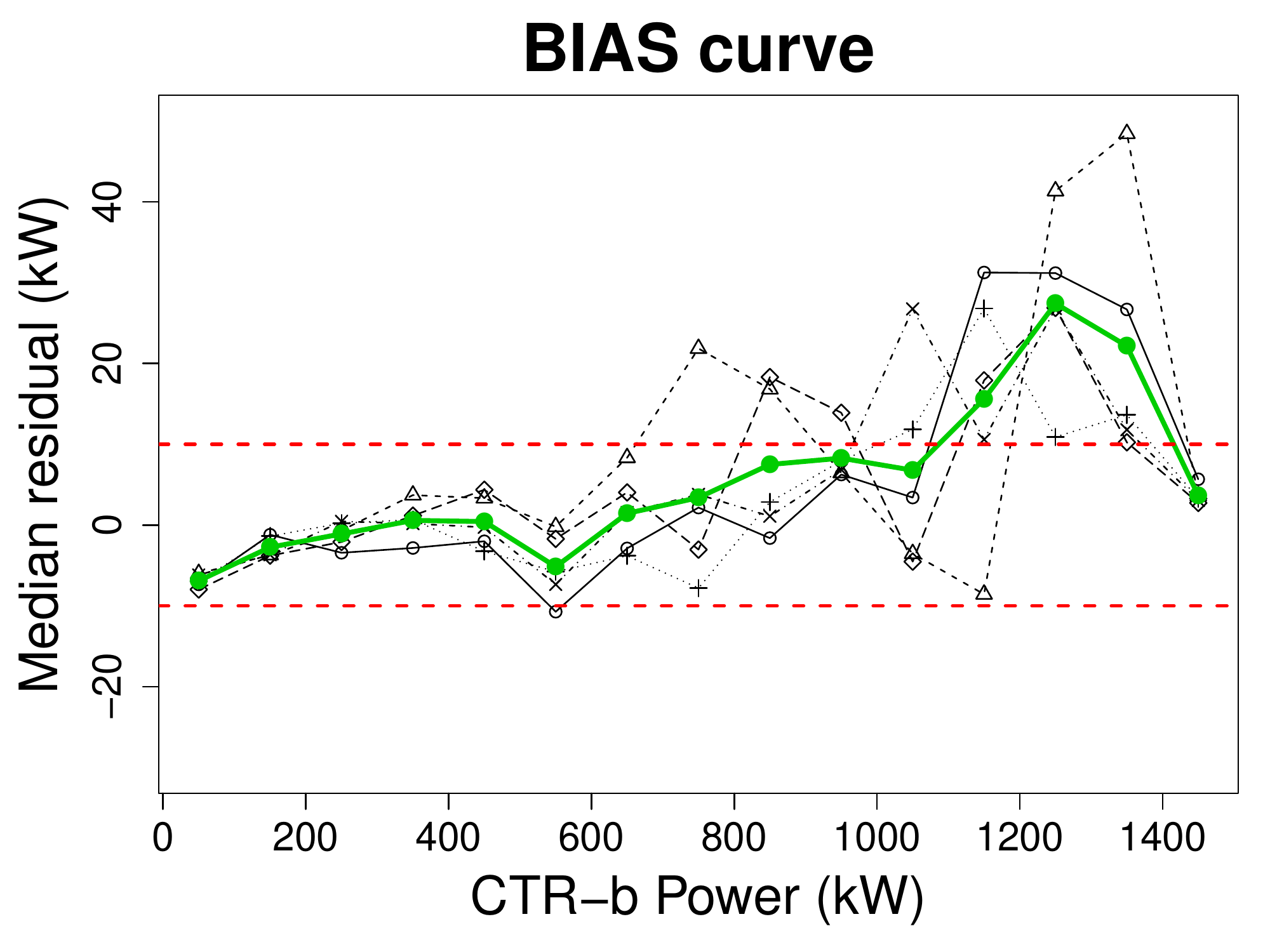}
	\caption{BIAS curve for a test turbine: the black lines denote $\text{BIAS}_b$ calculated for each of 5-fold CV, and the green line indicates their averages, i.e., $\text{CV}_{\text{BIAS}_b}$.}
	\label{fg:biasCurv}
\end{figure}

We use three performance measures to assess the prediction from a test turbine model and a control turbine model. The three measures include root mean square error (RMSE), empirical bias (BIAS), and residuals binned by control turbine power (BIAS curve). Suppose that $(\bold{x}_i, y_i)$ for $i=1,\ldots,n_{TR}$ denotes training data and $(\bold{x}_j, y_j)$ for $j=1,\ldots,n_{TS}$ denotes test data. Then,
\begin{equation}\label{eq:errs}
\begin{split}
\text{RMSE} &= \sqrt{\frac{1}{n_{TS}} \sum_{j=1}^{n_{TS}}{\left( y_j - \hat{f}(\bold{x}_j) \right)^2}}, \quad \text{and} \\
\text{BIAS}(\%) &= \frac{\sum_{j=1}^{n_{TS}}{(\hat{f}(\bold{x}_j) - y_j)}}{\sum_{j=1}^{n_{TS}}{\hat{f}(\bold{x}_j)}}.
\end{split}
\end{equation}
To generate the BIAS curve, we first bin the test data according to control turbine power by using the bin size of 100kW. Suppose that there are $m$ bins in total and $B_b$ for $b=1,\ldots,m$ is a set of test data points that belong to the $b$th bin. We define the bin-wise BIAS curve as
\begin{equation}
\text{BIAS}_b = \underset{j \in B_b}{\text{median}} (y_j - \hat{f}(\bold{x}_j)), \quad b = 1, \ldots, m.
\label{eq:biasCurv}
\end{equation}
Then, the BIAS curve can be generated by $\text{BIAS}_b$ and the mid-point of each power bin $b$ for $\forall b = 1, \ldots, m$; see Figure~\ref{fg:biasCurv}.

The three measures can be calculated for each data split. For the evaluation of a model, we use the averages of the three measures over five splits, referred to, respectively, as $\text{CV}_{\text{RMSE}}$, $\text{CV}_{\text{BIAS}}$, and $\text{CV}_{\text{BIAS}_b}$ in the subsequent sections.

\subsection{Variable selection}\label{ssc:varSel}
Wind speed is the most important variable determining wind power output. Typically, wind speed is measured at an anemometer on the nacelle of a WTG located behind the rotor. As passive device installation, especially VG installation, changes aerodynamic behaviors behind the rotor, the after-rotor measurement of wind speed for a same inflow wind can be different before and after the device installation. Therefore, using wind speed measured at a test turbine, albeit the most influential, would make the extraction of device installation effect on power more challenging.

One way to overcome the inconsistent measurement problem is to use the measurements of a nearby WTG for which there is no change of the after-rotor measurements. Recall that we would use a control turbine for calibration purpose. If we use wind speed measured at this control turbine, the control turbine model will take advantage of using its own measurements yielding relatively small prediction error. However, such an advantage is not applicable to a test turbine model, eventually resulting in a larger difference of the prediction error between the two models.

Due to all these considerations, a different source of wind speed measurements is needed, and the source can be another nearby WTG or a met mast. In general, one can find a WTG that is closer to a test turbine than a mast is.  Hence, we choose another control turbine as a source of wind speed measurements; this turbine is referred to as a neutral turbine.

Let REF, CTR-b, and CTR-n denote the test WTG, the baseline control WTG for calibration, and the neutral control WTG providing the measurements of some key covariates like wind speed. Then, we consider the following variables for modeling a multi-dimensional power response surface:
\begin{itemize}\setlength\itemsep{0pt}
	\item $V$-CTRn: wind speed measured at the neutral turbine (CTR-n),
	\item $dV$-CTRn: wind speed difference between time $t$ and $t-1$, calculated from $V$-CTRn,
	\item $PW$-CTRn: power output of the neutral turbine (CTR-n),
	\item Direction: wind direction measured at the reference turbine (REF),
	\item Density: air density calculated by the temperature and air pressure measurements at the reference turbine (REF),
	\item Hour: hour of the day for each measurement.
\end{itemize}
We do not consider other variables measured at a met mast because, conditioned on the use of the above variables, prediction quality does not improve in terms of RMSE.  The datasets associated with the met mast have also a much high percentage of missing values.  If we have to use both nacelle data and mast data, the total amount of data in fact become less after data alignment.

To determine the best set of variables, we first start with a model having the full set of variables (saying currently optimal) and calculate its $\text{CV}_{\text{RMSE}}$. Then, we eliminate a variable at a time constructing multiple $(q-1)$-variable models, where $q$ is the number of variables in the current optimal model. We find a $(q-1)$-variable model with the least $\text{CV}_{\text{RMSE}}$, and if its $\text{CV}_{\text{RMSE}}$ is lower than that of the current optimal model, the $(q-1)$-variable model becomes a new current optimal model. We continue this while decreasing $q$ by one until there is no reduction in $\text{CV}_{\text{RMSE}}$; in this case, the current optimal model becomes the optimal model. The optimal set of variables will be denoted by $\bold{x}^*$ for subsequent analysis. Algorithm~\ref{alg:varSel} shows the details of the variable selection process.

\begin{algorithm}[h!]
\caption{Selection of an optimal set of variables}\label{alg:varSel}
	Initialize $q \gets p$, $\mathcal{S} \gets \{x_1, x_2, \ldots, x_p \}$\;
	For each of five training/testing splits, predict $\hat{f}(\bold{x}_j)$ for $j=1,\ldots,n_{TS}$ by using $\mathcal{S}$ and calculate $CV_{RMSE}$\;
	$T \gets CV_{RMSE}$\;
	\Repeat{
		$l \gets 1$\;
		\While{$l \leq q$}{
			$\mathcal{S}^\prime \gets \mathcal{S}\setminus\{\text{the } l\text{th element of }\mathcal{S}\}$\;
			For each of five training/testing splits, predict $\hat{f}(\bold{x}_j)$ for $j=1,\ldots,n_{TS}$ by using $\mathcal{S}^\prime$ and calculate $CV_{RMSE}$\;
			$CV_l \gets CV_{RMSE}$\;
			$l\gets l+1$\;
		}
		$l^* \gets \argmin_{l \in \{1,\ldots, q\}}{CV_l}$\;
		\uIf{$CV_{l^*} < T$}{
			$q \gets q-1$\;
			$\mathcal{S} \gets \mathcal{S}^\prime$\;
			$T \gets CV_{l^*}$\;
		}
		\Else{
			\textbf{break}\;
		}
	}
	\Return{$\bold{x}^* \gets \mathcal{S}$}\;
\end{algorithm}

\subsection{Control turbine selection}\label{ssc:ctr}
As discussed in Section~\ref{ssc:varSel}, the quantification of potential power gain will use two types of control turbines, CTR-b and CTR-n. Both control turbines need to be determined based on the outcome of Period 1 analysis. The selection of CTR-b is more restrictive because, to satisfy the assumption of having the same unknown or unobservable effect with REF, it should typically be in a closer vicinity of REF than CTR-n. Sometimes, the selection of CTR-b can be obvious based on the proximity to REF, but depending on the layout of a wind farm, such an immediate selection may not be applicable in other cases.

\begin{figure}[t]
	\centering
	\includegraphics[width=0.6\textwidth]{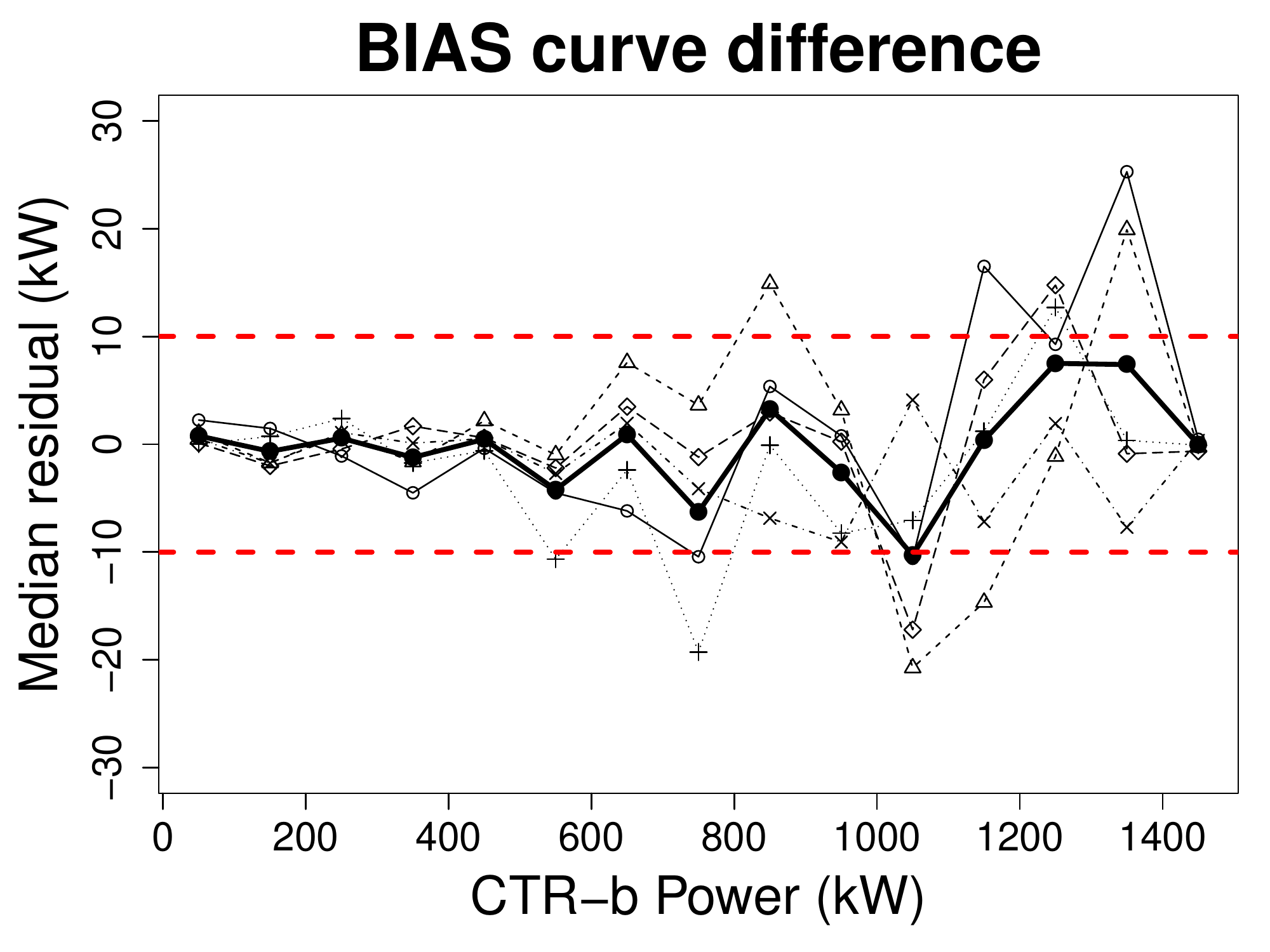}
	\caption{BIAS curve difference: the dotted lines denote $\text{DIFF}_b$ calculated for each of 5-fold CV, and the solid line indicates their averages, i.e., $\text{CV}_{\text{DIFF}_b}$.}
	\label{fg:curvdiff}
\end{figure}

For the selection of control turbines, we use REF model's $\text{CV}_{\text{RMSE}}$ and $\text{CV}_{\text{BIAS}}$ and also BIAS curve difference between REF model and CTR-b model defined by
\begin{equation}
\text{DIFF}_b = \text{BIAS}_b^\text{REF} - \text{BIAS}_b^\text{CTR-b}, \quad b=1,\ldots,m
\end{equation}
where $\text{BIAS}_b^\text{REF}$ and $\text{BIAS}_b^\text{CTR-b}$ are bin-wise BIAS curve for REF and CTR-b models, respectively. Similar to the notations of other performance measures, we use $\text{CV}_{\text{DIFF}_b}$ to denote the average of $\text{DIFF}_b$ obtained from five different data splits. An example of the BIAS curve difference is shown in Figure~\ref{fg:curvdiff}.

The BIAS curve difference, $\text{CV}_{\text{DIFF}_b}$, assesses the similarity of unexplained variation between REF and CTR-b models while $\text{CV}_{\text{RMSE}}$ and $\text{CV}_{\text{BIAS}}$ evaluate the prediction accuracy of REF model. Considering that the unexplained variation of REF model is corrected by that of CTR-b model, the BIAS curve difference would be the most important criterion justifying the assumption for calibration. Still, $\text{CV}_{\text{RMSE}}$ and $\text{CV}_{\text{BIAS}}$ values need to be kept at a sufficiently low level to minimize the amount of correction.

As the selection becomes a multi-criteria decision, we do not impose an absolute threshold for each performance measure. In general, there is a limited number of WTGs in the vicinity of REF, so we rather find the best pair of CTR-b/CTR-n among all possible alternatives. Yet, a good pair needs to have
\[|\text{CV}_{\text{DIFF}_b}| \leq 10 \text{kW} \quad \text{for} \quad b=1,\ldots,m \]
or at least close to the limit of 10kW as well as a relatively low $\text{CV}_{\text{RMSE}}$ and $\text{CV}_{\text{BIAS}}$, as compared to other pairs.

\section{Period 2 analysis}\label{sc:per2}
In Period 2 analysis, we use the best set of variables, $\bold{x}^*$, and CTR-b and \\CTR-n determined in Period 1 analysis to quantify the potential gain of passive device installation. For the gain quantification, we evaluate how the change of unexplained variation over Period~1 and Period~2 is different between REF and CTR-b while developing a machine learning model for each of them that shares the same $\bold{x}^*$.

\begin{figure}[t!]
	\centering
	\includegraphics[width=\textwidth]{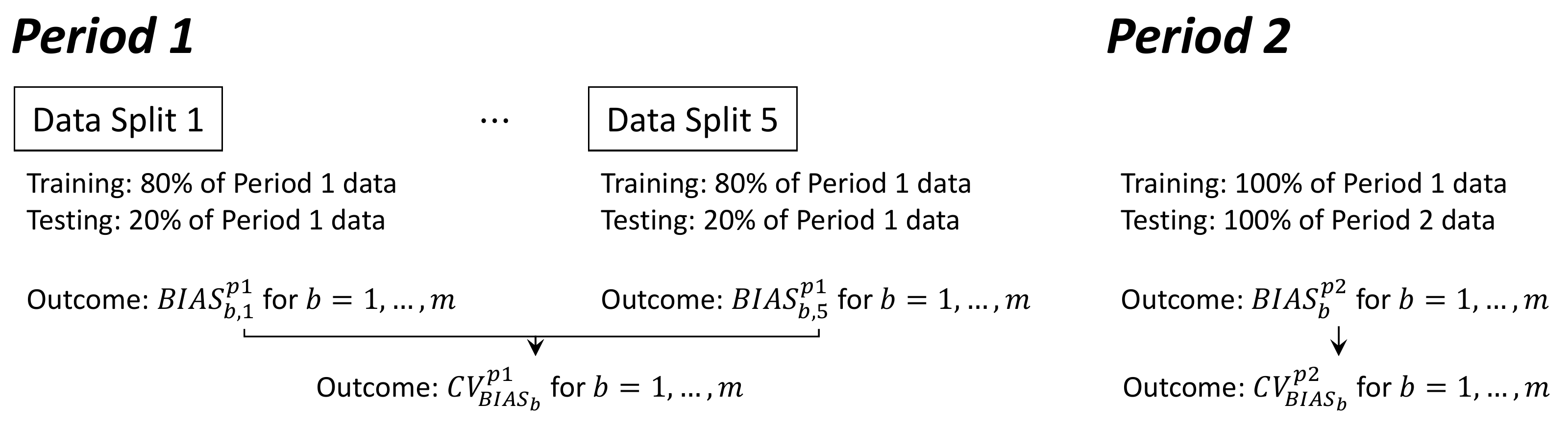}
	\caption{Data structure for gain quantification}
	\label{fg:dataP2}
\end{figure}

As the quantification requires to control for unexplained variation for both Period~1 and~2, we need prediction for both periods. For an accurate estimation of unexplained variation, we calculate out-of-sample prediction error while keeping the 5-fold CV structure for the Period~1 prediction. The Period~2 prediction does not require data splitting since the entire Period~1 data can be used to train a model.  Then, the out-of-sample prediction error can be calculated from Period~2 data. Figure~\ref{fg:dataP2} illustrates data structure for Period~2 analysis.

From the Period~1 prediction, we have five $\text{BIAS}_{b}^{p1}$ for each data split per each bin $b$, where the superscript $p1$ indicates that the value is resulting from the Period~1 prediction. By taking the average of five $\text{BIAS}_{b}^{p1}$, we calculate $\text{CV}_{\text{BIAS}_b}^{p1}$ for $b=1,\ldots,m$. On the other hand, the Period~2 prediction provides a single $\text{BIAS}_{b}^{p2}$ for each bin, so we define $\text{CV}_{\text{BIAS}_b}^{p2} = \text{BIAS}_{b}^{p2}$ for $b=1,\ldots,m$ without the need to average.

As we have REF model and CTR-b model, $\text{CV}_{\text{BIAS}_b}^{p1}$ and $\text{CV}_{\text{BIAS}_b}^{p2}$ can be calculated for each model. Let superscript REF and CTR-b indicate that the values are derived from each of the two models, respectively. For example, $\text{CV}_{\text{BIAS}_b}^{\text{REF},p1}$ represents the average BIAS of REF model in Period~1, and $\text{CV}_{\text{BIAS}_b}^{\text{CTR-b},p2}$ denotes the BIAS of CTR-b model in Period~2.

\begin{figure}[t!]
		\centering
		\begin{subfigure}{0.48\textwidth}
			\centering
			\includegraphics[width=\textwidth]{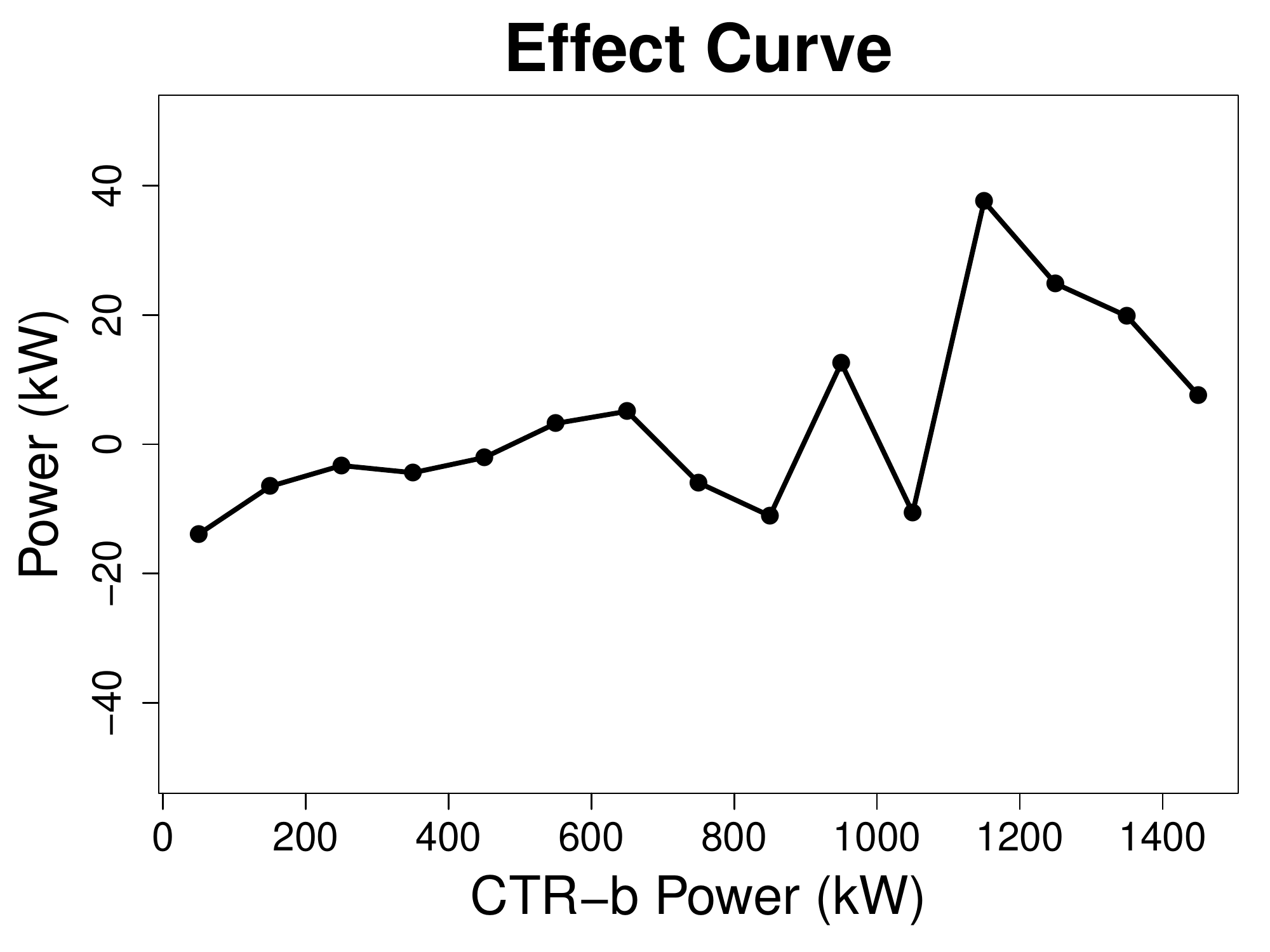}
		\end{subfigure}
		\begin{subfigure}{0.48\textwidth}
			\centering
			\includegraphics[width=\textwidth]{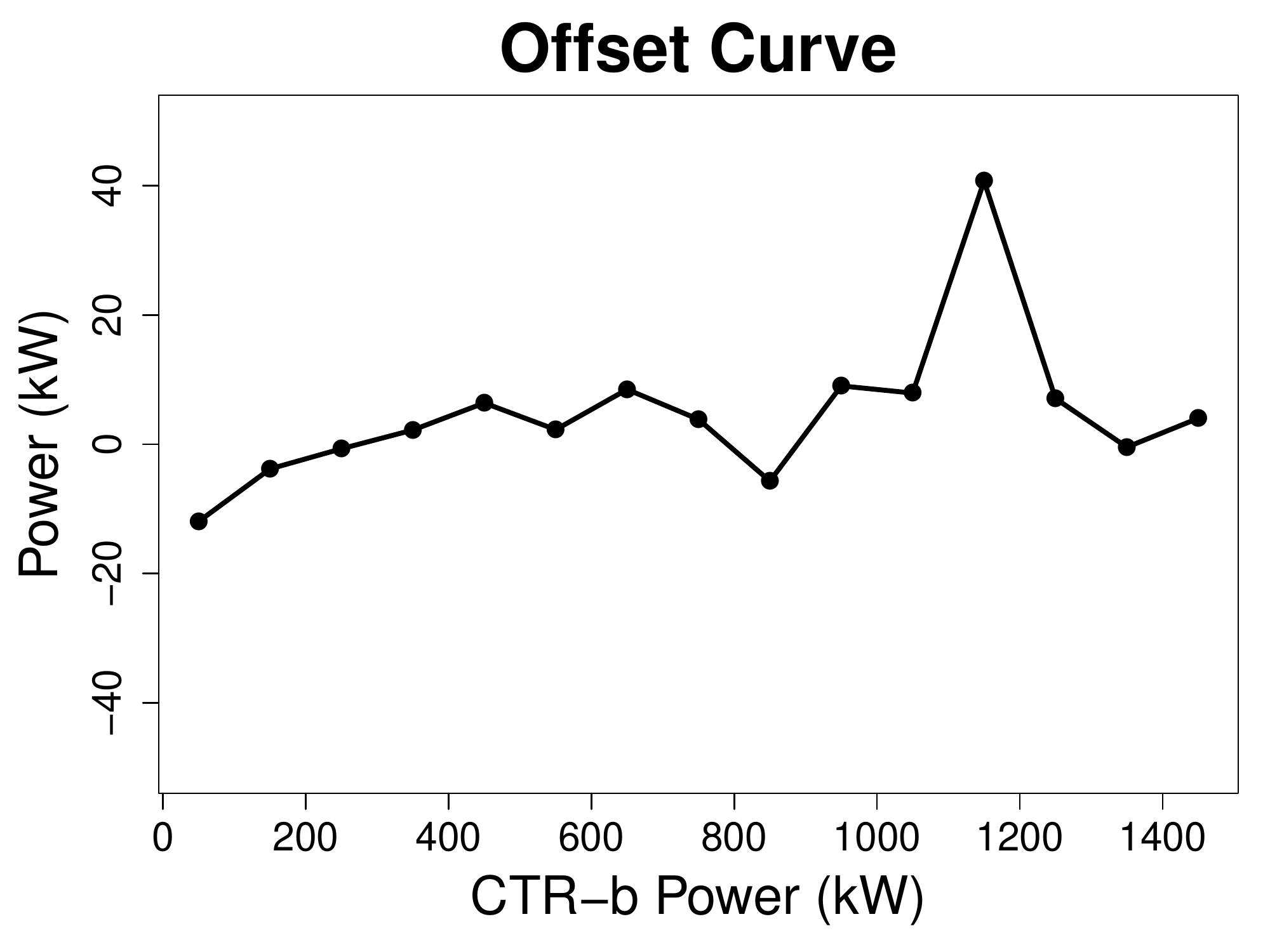}
		\end{subfigure}		
		\caption{Curves illustrating temporal change in power not explained by given covariates, $\bold{x}$: (a) Effect curve and (b) Offset curve.}
		\label{fg:effoff}
\end{figure}

We define Effect curve as the change in the BIAS curve of REF power over Period~1 and~2, expressed by
\begin{equation}
\text{Effect}_b = \text{CV}_{\text{BIAS}_b}^{\text{REF},p2} - \text{CV}_{\text{BIAS}_b}^{\text{REF},p1}.
\end{equation}
Since passive devices are installed on the REF turbine in Period~2, $\text{Effect}_b$ includes the change in REF power due to passive device installation and the change in REF power due to unknown factors (other than known $\bold{x}$). If we define a similar quantity for CTR-b model, referred to as Offset curve, then
\begin{equation}
\text{Offset}_b = \text{CV}_{\text{BIAS}_b}^{\text{CTR-b},p2} - \text{CV}_{\text{BIAS}_b}^{\text{CTR-b},p1}.
\end{equation}
CTR-b turbine is without any passive device installation, even in Period~2, so $\text{Offset}_b$ explains the change in CTR-b power due to unknown factors other than $\bold{x}$. We assumed that this change caused by unknown factors is the same for REF turbine and CTR-b turbine, so by taking the difference of $\text{Effect}_b$ and $\text{Offset}_b$, we quantify the (bin-wise) change in REF power due to passive device installation as follows:
\begin{equation}
\text{Gain}_b = \text{Effect}_b - \text{Offset}_b, \quad b= 1, \ldots, m.
\end{equation}
The examples of Effect curve and Offset curve are shown in Figure~\ref{fg:effoff}.

To aggregate the bin-wise gain defined by Gain curve into a single quantity, we calculate a weighted sum of $\text{Gain}_b$. For the weights, we use the long-term annual power frequency, expressed in the unit of hours and estimated using the long-term historical data. Let $\pi_b$ denote the annual power frequency of bin $b$. Then, annualized Gain is finally estimated by
\begin{equation}
\text{Gain}(\%) = \frac{\sum_{b=1}^m{\pi_b(\text{h}) \cdot \text{Gain}_b \,\, (\text{kW})}}{\text{AEP} \,\, (\text{kWh})},
\label{eq:gain}
\end{equation}
where AEP is also estimated from long-term historical data and OEM power curve.

To reduce the uncertainty of the gain quantification, we can apply bootstrap and establish a confidence interval of annualized Gain. Bootstrapping relies on random sampling with replacement. Suppose that $\mathcal{Z} = \{(\bold{x}_i, y_i): i=1,\ldots,n \}$ includes all given data points (both Period~1 and~2). We draw $n$ random samples from the original dataset, $\mathcal{Z}$, while allowing that a single data point is sampled multiple times. These sampled data points form a bootstrapped dataset of which sample size is $n$. By repeating this $B$ times, we can construct $\mathcal{Z}_g$ for $g=1,\ldots,B$ where $\mathcal{Z}_g$ is a bootstrapped dataset and $B$ is the total number of bootstrapping. For each $\mathcal{Z}_g$, we partition the bootstrapped dataset into Period~1 and Period~2 data based on time stamps and calculate a single value of annualized Gain following Eq.~\eqref{eq:gain}; note, for this calculation, prediction is always based on the optimal set of variables determined in Period~1 analysis. Finally, we have annualized Gain quantified $B$ times, and these quantified values are used to build a confidence interval of annualized Gain. For example, if $B=10$, by eliminating the lowest annualized Gain and the greatest annualized Gain and by taking the range of the rest eight values, we can have an 80\% confidence interval (eight out of ten).

\section{References}\label{ref}
\begin{enumerate}
\item Lee, Ding, Genton and Xie (2015a) ``Power curve estimation with multivariate environmental factors for inland and offshore wind farms," \emph{Journal of the American Statistical Association}, \textbf{110}: 56-67.

\item Lee, Ding, Xie and Genton (2015b) ``Kernel Plus method for quantifying wind turbine upgrades," \emph{Wind Energy}, \textbf{18}: 1207-1219.
    
\item Hwangbo, Ding, Eisele, Weinzierl, Lang and Pechlivanoglou (2017) "Quantifying the effect of vortex generator installation on wind power production: an academia-industry case study," \emph{Renewable Energy}, \textbf{113}: 1589-1597.
    
\item Shin, Ding, and Huang (2018). "Covariate matching methods for testing and quantifying wind turbine upgrades," \emph{The Annals of Applied Statistics}, \textbf{12}: 1271-1292.

\item Yu Ding (2019) \emph{Data Science for Wind Energy}, Boca Raton, FL: Chapman \& Hall/CRC Press.
\end{enumerate}
\end{document}